# Transimpedance Amplifier with Automatic Gain Control Based on Memristors for Optical Signal Acquisition

Sariel Hodisan, *Student Member, IEEE,* and Shahar Kvatinsky, *Senior Member, IEEE*

**Abstract**

Transimpedance amplifiers (TIA) play a crucial role in various electronic systems, especially in optical signal acquisition. However, their performance is often hampered by saturation issues due to high input currents, leading to prolonged recovery times. This paper addresses this challenge by introducing a novel approach utilizing a memristive automatic gain control (AGC) to adjust the TIA's gain and enhance its dynamic range. We replace the typical feedback resistor of a TIA with a valence-change mechanism (VCM) memristor. This substitution enables the TIA to adapt to a broader range of input signals, leveraging the substantial OFF/ON resistance ratio of the memristor. This paper also presents the reading and resetting sub-circuits essential for monitoring and controling the memristor's state. The proposed circuit is evaluated through SPICE simulations. Furthermore, we extend our evaluation to practical testing using a printed circuit board (PCB) integrating the TIA and memristor. We show a remarkable 40 dB increase in the dynamic range of our TIA memristor circuit compared to traditional resistor-based TIAs.

**Index Terms**

Transimpedance amplifier, wide dynamic range, AGC, optical receivers, memristor

## I. Introduction

TRANS-IMPEDANCE AMPLIFIERS (TIAs) are used in many optical and radio-frequency (RF) systems for communication, measurement, and bio-sensing applications [1]–[5]. TIAs convert the input current from a photo-diode or a previous $G_m$ stage into an amplified voltage signal. Since TIAs are the first (or one of the first) circuit blocks in receiving and sensing electronics circuits, they must provide high gain and low noise. By that, they can overcome the noise of the succeeding stages and maximize the total circuit signal-to-noise ratio (SNR) and sensitivity [6]. Designing a high dynamic range TIA for optical sensing applications is challenging since the photo-diode current has a high dynamic range. Injecting a high input current into the TIA stage can easily saturate the TIA and cause it to be insensitive to input signals for a while (blind time for the receiver circuit) [7]. Overcoming this limitation requires installing an automatic gain control (AGC) amplifier or other advanced techniques to create a wide dynamic range TIA that can withstand high input currents before being saturated. Section II explores several established techniques for achieving an extended dynamic range. Nevertheless, this improvement may entail trade-offs, such as diminished bandwidth, stability, linearity, power efficiency, and a larger circuit footprint.

In this paper, we present a novel method to design an AGC mechanism for the TIA based on emerging memristor technology. We use resistive random access memory (ReRAM) as the memristive technology. ReRAM performs resistive switching between a high resistance state (HRS or $R_{off}$ ) to a low resistance state (LRS or $R_{on}$). Applying an external short voltage pulse across the ReRAM memristor device activates the transition of the device from one resistive state to the other. Since both these states retain their values after the applied voltage is removed, ReRAM is a non-volatile memory device. ReRAM memristors exhibit a gradual RESET transition ($R_{on}$ to $R_{off}$ transition) and abrupt SET transition ($R_{off}$ to $R_{on}$ transition) [8], [9]. This feature makes them eminently suitable for high-speed switching to a lower resistance once the voltage applied to them reaches a threshold voltage. We focus on the valence-change mechanism (VCM) ReRAM memristor [10] because of its abrupt switching to an $R_{on}$, high endurance, and moderate threshold voltages.

Figure 1 shows the proposed solution for increasing the TIA stage dynamic range using a memristor device. The incorporated memristor grants the TIA AGC capability. The memristor is located in the operational amplifier's feedback to set the gain of the TIA. Since its resistance can be switched when the current through the memristor varies (and accordingly, there is a voltage drop across it), the memristor is used to provide AGC capabilities to the TIA. We use this memristor to switch the gain of the TIA between HRS and LRS. The gain diminution happens automatically and very fast as the current through the TIA stage increases and passes a certain threshold before the TIA saturates. This method extends the TIA's dynamic range in line with the memristor's $\frac{R_{off}}{R_{on}}$ ratio. Our proposed design includes a reading circuit that monitors the memristor state. Once

This work was partially funded by the Israel Innovation Authority through the Promoting Applied Research in Academia Program, grant number 81755.

Sariel Hodisan and Shahar Kvatinsky are with the Andrew and Erna Finci Viterbi Faculty of Electrical and Computer Engineering, Technion – Israel Institute of Technology, Haifa 3200003, Israel (e-mail: hsariel@campus.technion.ac.il; shahar@ee.technion.ac.il).

Sariel Hodisan is also with Applied Materials Corporation.

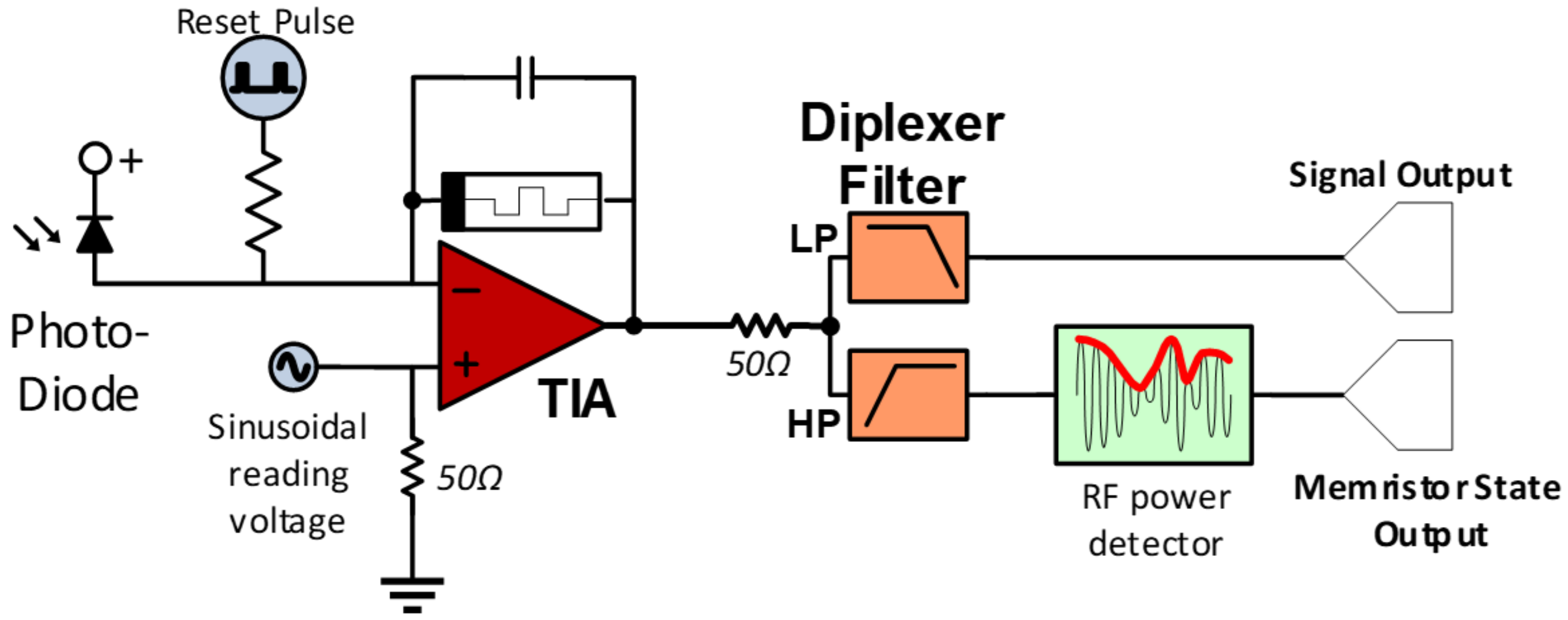

Fig. 1. Memristor-based transimpedance AGC amplifier. A photo-diode is connected to a TIA based on an operational amplifier. The memristor is placed in the operational amplifier feedback path to provide its AGC capability.

the TIA gain needs to be changed back to high impedance (to allow it to receive low amplitude currents), our circuit allows a reset pulse to be applied to the operational amplifier input to restore the memristor to a high resistance state.

We first confirm the validity of our proposed circuit using SPICE simulations. Following this, we experimentally demonstrate our design using a full board containing the TIA stage, a $Pt/W/Ta_2O_5/Pt/Ti$ VCM memristor, and auxiliary circuits. Our experimental results show good agreement with the circuit behavior seen in simulations. The measurements confirm that our solution enables the extension of the TIA's dynamic range by the memristor's $\frac{R_{off}}{R_{on}}$ ratio automatically, according to the input current signal level. For the memristor device we used, the dynamic range can be extended by approximately 40dB compared to resistor-based TIAs.

The rest of the paper is organized as follows. Section II presents the operation of high dynamic range TIA based on operational amplifiers. In section III, we detail the information regarding the used VCM memristor. Section IV presents a detailed explanation of the proposed circuit. Section V presents the manufactured board and its measured performance, along with a discussion of the limitations of the proposed solution. Section VI compares between this study and prior high dynamic range TIA designs. We conclude the paper in Section VII and discuss future work.

## II. High Dynamic Range Transimpedance Amplifiers

Transimpedance amplifiers are usually based on operational amplifiers with a feedback network [11]. A simplified block diagram of a basic TIA is shown in Figure 2. A resistor $R_F$ is placed in a negative feedback loop around an amplifier having an open loop gain $A$. As long as the open-loop gain of the amplifier is sufficiently high, the transimpedance gain of this circuit would be approximately $-R_F$ . This circuit is well-suited for direct connection to a current source due to its low input impedance.

Usually, a feedback capacitor, $C_F$ , is connected in parallel to the resistor $R_F$ to limit the amplifier bandwidth and improve stability. When a feedback capacitor is added, the $3dB$ bandwidth of the TIA is limited and becomes [6]

$$BW_{3dB} < \frac{\sqrt{2}}{2\pi R_F C_F}. \tag{1}$$

From (1), the feedback resistor $R_F$ controls the TIA gain and the $3dB$ bandwidth. It also controls the TIA's dynamic range. The dynamic range of the TIA (with SNR=1) can be characterized as the ratio between the maximum input current and the integrated equivalent input-referred current noise $I_n$:

$$Dynamic\ range = \frac{max(I_{in})}{\overline{I_n^2}}. \tag{2}$$

The feedback resistance, $R_F$ , serves as both a noise source for the TIA stage and a factor in determining the TIA's integrated equivalent input-referred current noise. This determination involves dividing the integrated output voltage noise of the TIA by $R_F$ , and the input-referred current noise is therefore.

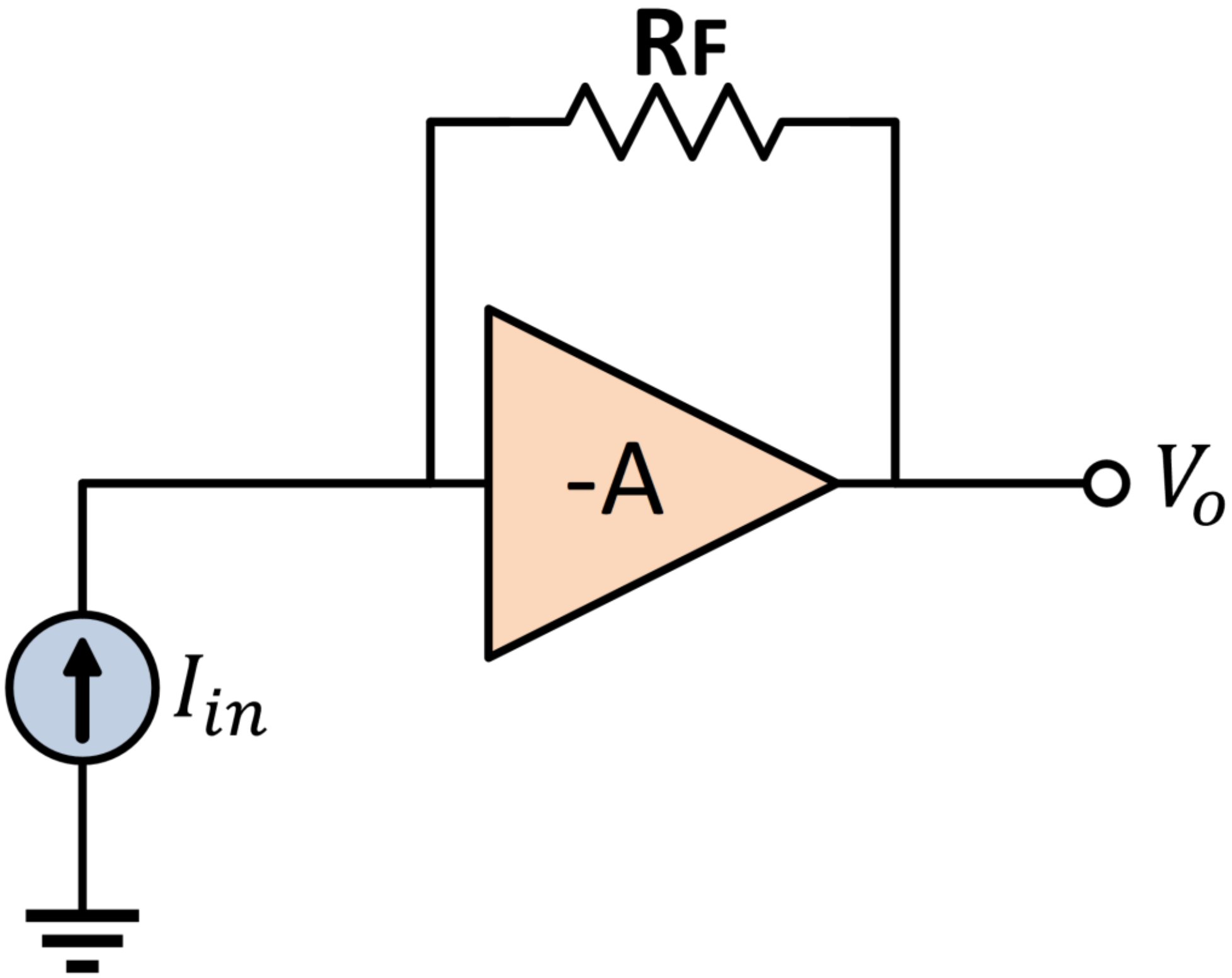

Fig. 2. Basic TIA circuit. $V_o = -I_{in}R_F$.

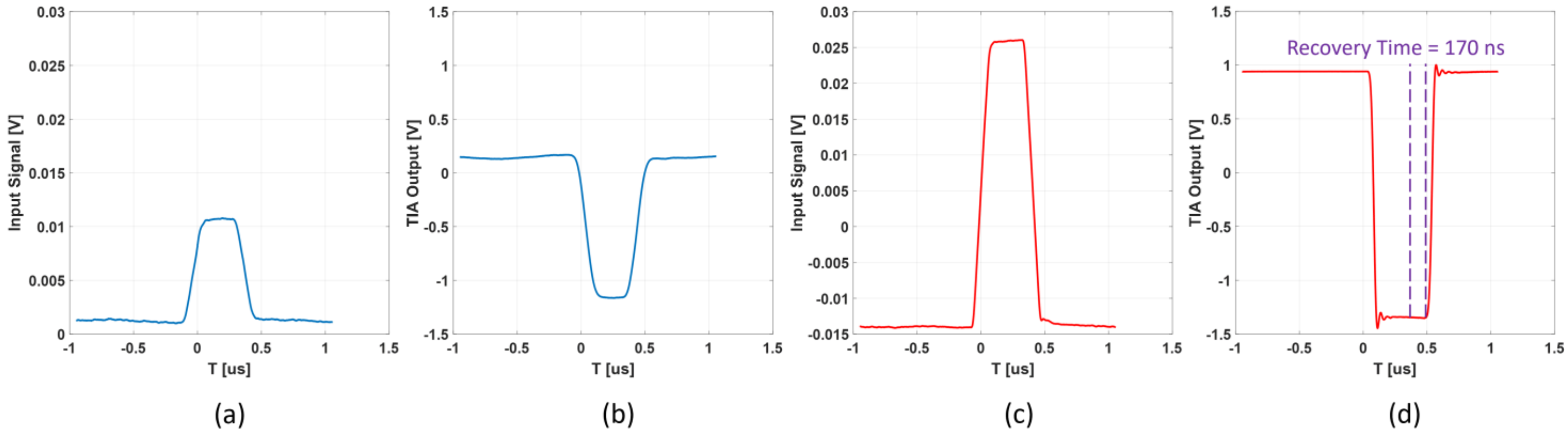

Fig. 3. TIA (based on the LMH6601 amplifier [12]) before and after saturation. (a) Small non-saturating input signal, and (b) the TIA output. (c) High input signal that saturates the TIA and (d) the saturated TIA output. There is a considerable time delay until the TIA recovers and can respond to the input signal.

$$\overline{I_n^{\ 2}} = \frac{\overline{V_{no}^{\ 2}}}{R_F}. \tag{3}$$

Where $V_{no}$ is the output noise of the TIA.

Using a high feedback resistor value, $R_F$, is beneficial to decrease the input-referred current noise. However, a high feedback resistor value can limit the maximum input current that the TIA stage can amplify, as the amplifier's output voltage is limited by its supply voltage. The larger the feedback resistor value is, the lower the maximum input current the TIA can amplify, and therefore

$$max(I_{in}) = \frac{max(V_o)}{R_F}. \tag{4}$$

where $V_o$ is the output voltage of the amplifier.

From (2)–(4), the feedback resistor value must be carefully selected to maximize the TIA's dynamic range. Furthermore, since the feedback resistor, $R_F$ , also sets the TIA stage gain and bandwidth, the TIA's gain trades off the bandwidth and dynamic range.

When the input current exceeds the maximum current of the TIA stage, it detrimentally affects the amplifier's performance. Firstly, the TIA becomes incapable of amplifying the input signal, as it is designed to operate within a specific current range. Amplifier saturation occurs, where the output voltage is stuck at its extreme value, rendering the amplifier unresponsive to variations in the input current. This insensitivity persists for an undefined duration as long as the input current remains beyond the specified limits, making the TIA non-functional for signal processing. Continuously exceeding the maximum input current rating can cause long-term damage to the amplifier. The excessive current can lead to increased power dissipation, which may result in overheating and, ultimately, damage to the amplifier's internal components. This can permanently impair the TIA's performance or even render it non-functional. Figure 3 illustrates this saturation phenomenon.

Several methods have been proposed to untangle the above tradeoffs and increase the TIA dynamic range. One method is to divide the TIA gain between two or more stages and have voltage amplifiers as the subsequent stages [13]–[17]. This method allows the TIA to have a lower feedback resistance value (gain) and become saturated for higher input currents. Still, it increases the total noise, power, and area of the whole circuit. Another method uses a nonlinear feedback network containing a diode or a transistor to implement a logarithmic TIA [18]. Such a TIA can operate over several decades of input current, substantially increasing the TIA's input overload current [19]. The drawbacks of this method are increased noise, amplifier instability, and nonlinearity for large signals.

The dynamic range can also be increased by reducing the noise of the TIA stage. This allows the TIA to have a lower gain (lower feedback resistor value) as less gain is needed to overcome the noise of the following stages. TIA noise can be reduced by using lower noise amplifiers (improved transistor process technology) or by lowering the TIA input capacitance using a common-base (or gate) stage before the TIA [20]. Implementing such a method necessitates meticulous trade-off design, as the contribution of the common-base stage to the overall noise can offset the reduction achieved in TIA noise. Another solution when the dynamic range of the photodiode is very high, and high input currents are to be expected, is to split the high current input among several parallel TIA stages [21]. The drawback of this method would be the extra power and area used for the design.

Another popular way to increase the TIA dynamic range is using an *automatic gain control* (AGC) mechanism [22]–[27]. An AGC is a control loop that measures the output voltage of the TIA. Accordingly, it controls the TIA gain by fine-tuning (or switching) its feedback resistor or a variable input current attenuator. Several commercial TIA devices using programmable gain are available [28], [29]. The drawbacks of this method are the extra circuit complexity needed for the control loop, the need to assure stability for all TIA possible gain values, and the non-instantaneous gain switching (or tuning) time. We propose a TIA with an AGC mechanism based on a memristor device to overcome these obstacles.

## III. ReRAM VCM

Resistive switching devices (ReRAMs) are among the most common memristor technologies. The resistive switching of ReRAM devices is based on the growth of a conductive filament inside the device's dielectric due to the application of an external voltage greater than $V_{SET}$ [30]. When this filament connects the top and bottom electrodes of the device, a low resistance state ($R_{on}$) is obtained. Once an inverse voltage (for a bipolar switching device) is applied across the device and exceeds the value of $V_{RESET}$ , the conducting filament is ruptured, and the device switches to a high resistance state ($R_{off}$ ). Depending on the material composition of the filament, ReRAM devices can be classified as one of two main types: electrochemical metallization (ECM) or valence change mechanism (VCM).

We choose a valence-change mechanism (VCM) memristor for this work due to its abrupt SET transition [8], [9]. Abrupt transitioning to a lower resistance would allow our proposed AGC mechanism to be activated almost instantly. A cross-sectional diagram and a microscope image of a pair of the VCM memristors are shown in Figure 4. The VCM memristor underwent testing and characterization using a probe station to assess its parameter values. The average values obtained were as follows: $R_{off} = 15k\Omega$, $R_{on} = 120\Omega$, $V_{SET} = 0.8V$ , and $V_{RESET} = 0.8V$ . Additionally, the expected cycle-to-cycle variance of these parameters was measured. Figure 5 shows the measured I-V curve of 200 switching cycles for the VCM memristor. The memristor's $\frac{R_{off}}{R_{on}}$ ratio is approximately 100 on average, sufficiently high to extend the dynamic range of the TIA by 40dB.

Based on the experimental results, the memristor was modeled in SPICE using a binarized state switching model [32]. This model easily configures the $R_{off}$ and $R_{on}$ values, switching speed, and threshold voltages. The model also exhibits abrupt SET transition. The disadvantage of this model is its lower accuracy in modeling the RESET transition. The fabricated VCM memristor is integrated with the TIA circuitry by mounting it on a printed card board (PCB) carrier. The memristor wafer was diced into pairs of dies and then glued to a PCB carrier. Following this, it was wired to the PCB pads, under a protective enclosure, as shown in Figure 6.

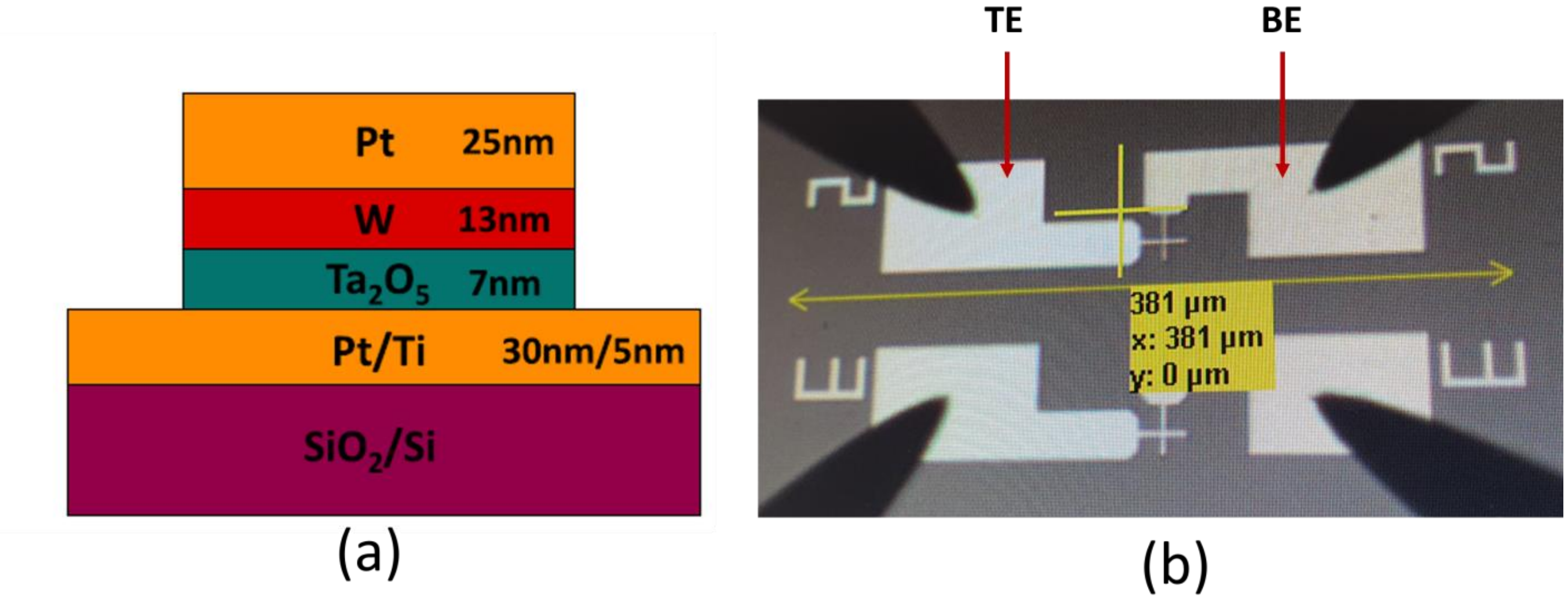

Fig. 4. (a) Cross-section diagram of a VCM memristor, reprinted from [31]. (b) Top view microscope image of a pair of VCM memristors

## IV. Memristor-based TIA architecture

### A. Proposed TIA

The proposed TIA works as follows. The memristor is initiated at a high resistive state (HRS); therefore, the TIA has a high gain as required for the system bandwidth and sensitivity. As long as the input current is of relatively low magnitude, the TIA operates as a regular TIA. When the input current increases and approaches the maximum current, the TIA can amplify before it saturates, the circuit outputs a sufficiently high voltage to SET the memristor. Since the top electrode (TE) of the memristor is held at virtual ground, the bottom electrode's voltage needs to be negative with an absolute value higher than the SET voltage of the memristor to switch it into a low resistance state. Once the memristor switches, the TIA gain is lowered. As long as the TIA gain is reduced, it can withstand high current levels without being saturated. Incorporating the memristor into the TIA's feedback network allows the circuit to gain AGC capability and increases the dynamic range by the memristor's $\frac{R_{off}}{R_{on}}$ ratio. Figure 7 illustrates the TIA operation. Upon reducing the TIA gain, there is also increase in bandwidth. As long as this lowered gain is employed to mitigate TIA saturation, the resulting amplifier bandwidth expansion should not introduce adverse effects to the circuit's performance. Naturally, the amplifier must be designed in a manner that precludes oscillations when the gain diminishes. We have confirmed in simulations that the TIA does not exhibit oscillations under each gain configuration.

### B. Testing Configuration

Figure 8 shows a detailed schematic of the proposed TIA circuit. For the validation and evaluation of the proposed design, we replaced the photo-diode with an operational transconductance amplifier (OTA). The OTA (also known as a $G_m$ stage) converts voltage input signals into current signals and drives the input current signal into the TIA stage.

The circuit is constructed as follows: The input voltage arrives from the test equipment, and an attenuator limits its amplitude. Since the OTA has a high transconductance gain, the input voltage has to be limited (otherwise, the OTA would be saturated easily). The OTA stage (based on the LT1228 amplifier [33]) converts the input voltage signals into current signals that flow to the TIA stage. Since high currents can damage the memristor, the maximum current is limited to a compliance current of $500\mu A$ using a common-base stage. The common-base stage acts as a current buffer (having a unity current gain) for a current lower than the compliance current. The current arriving at the TIA stage flows through the feedback memristor ($R_{mem}$) and is converted into voltage (using the LMH6601 [12] as the operational amplifier for the TIA stage). Since the voltage at the operational amplifier non-inverting input is at virtual ground, the output voltage at the TIA output is negative. The TIA output is

$$V_{TIA,out} = -I_{in} \cdot R_{mem}. \tag{5}$$

A $50\Omega$ resistor is introduced at the TIA output to ensure impedance matching, a necessary step given the closed-loop amplifier's low output impedance. The TIA output is connected to a Hi/Lo diplexer. The diplexer separates the amplified signal output from a low-pass filter branch and a high-frequency sinusoidal measurement signal that indicates the memristor's state. The high-pass filter branch is connected to an RF power detector. The RF detector converts the high-frequency sinusoidal signal to a DC voltage according to the amplitude of this signal.

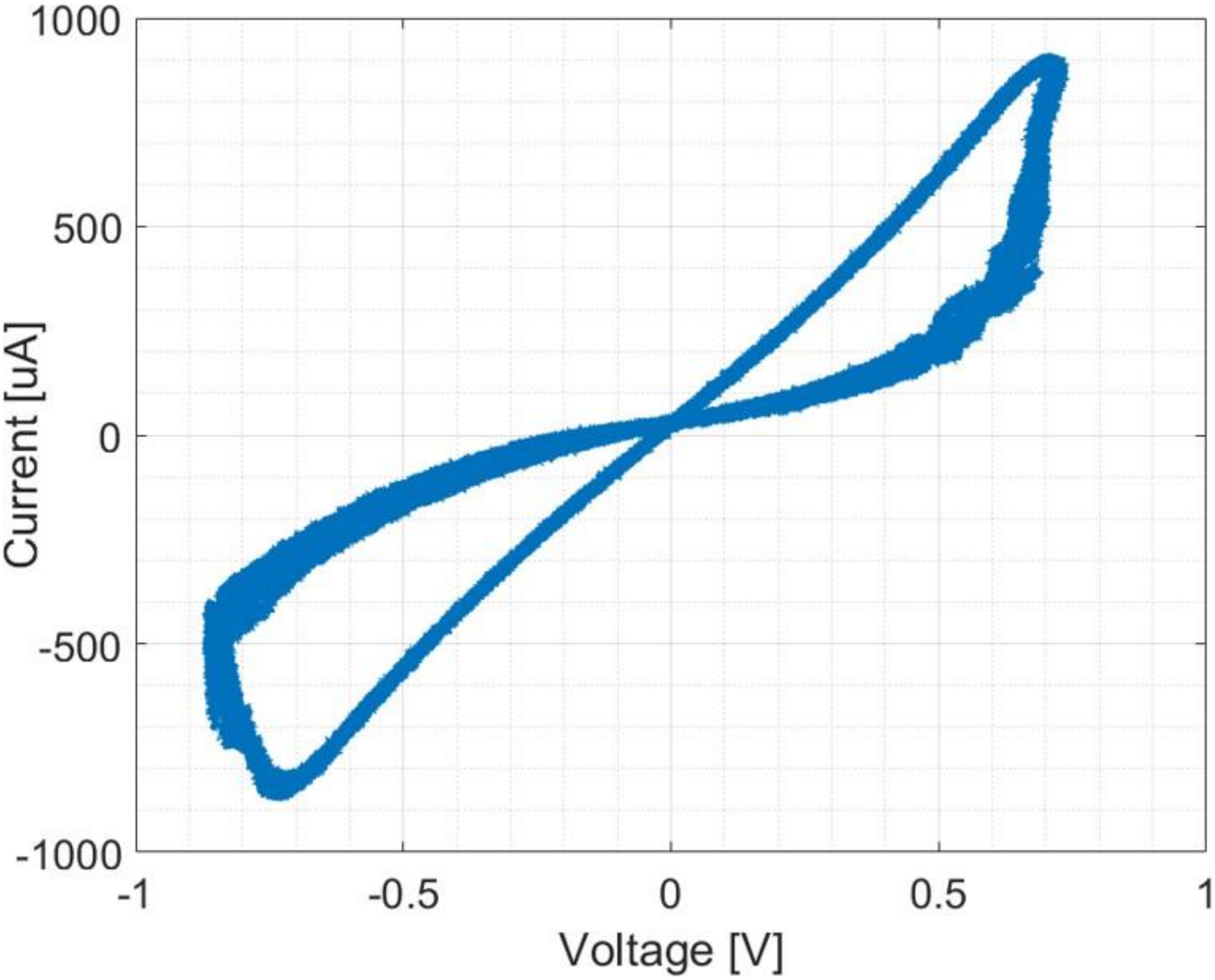

Fig. 5. VCM memristor I-V curve of 200 switching cycles. The measured values are on average $R_{off}$ = $15k\Omega$, $R_{on}$ = $120\Omega$, $V_{SET}$ = $0.8V$, and $V_{RESET} = -0.8V$.

### C. TIA Gain Monitoring

The memristor state has to be monitored to track the TIA gain. Tracking the TIA gain is necessary to know the correct input current levels and whether the memristor switched. This monitoring has to be performed without affecting the memristor state. We add reading circuits to monitor the memristor resistance state, as shown in Figure 9. The non-inverting input of the operational amplifier is connected to a sinusoidal source. The operational amplifier amplifies this sine wave (non-inverting configuration amplification). The amplification gain for the reading sine signal is

$$Reading\ signal\ gain = 1 + \frac{R_{mem}}{R_C}, \quad (6)$$

where $R_C$ is the resistor connected to the NPN transistor collector (see Figure 8). Therefore, the reading sine waveform is amplified according to the memristor resistance. This amplified waveform is combined with the TIA output signal.

The frequency of the sine wave signal being read is intentionally chosen to be higher than the TIA's bandwidth (typically two to three times higher). While this approach may present challenges for high-frequency TIAs operating in the GHz range, our ongoing research endeavors are dedicated to developing a solution that obviates the need for constant monitoring of the memristor's state. The Hi/Lo diplexer performs the crucial task of separating the reading signal from the TIA amplified signal. The TIA output is filtered through a low-pass filter to eliminate the sine wave component, while the sine wave is directed through a high-pass filter branch, amplified, and then conveyed to an RF power detector (LTC5507 [34]). The RF power detector output is a DC signal proportional to its input voltage's logarithm. The memristor resistance state is deduced by sampling and monitoring the RF power detector output level. Higher RF power detector output is translated into higher memristor resistance.

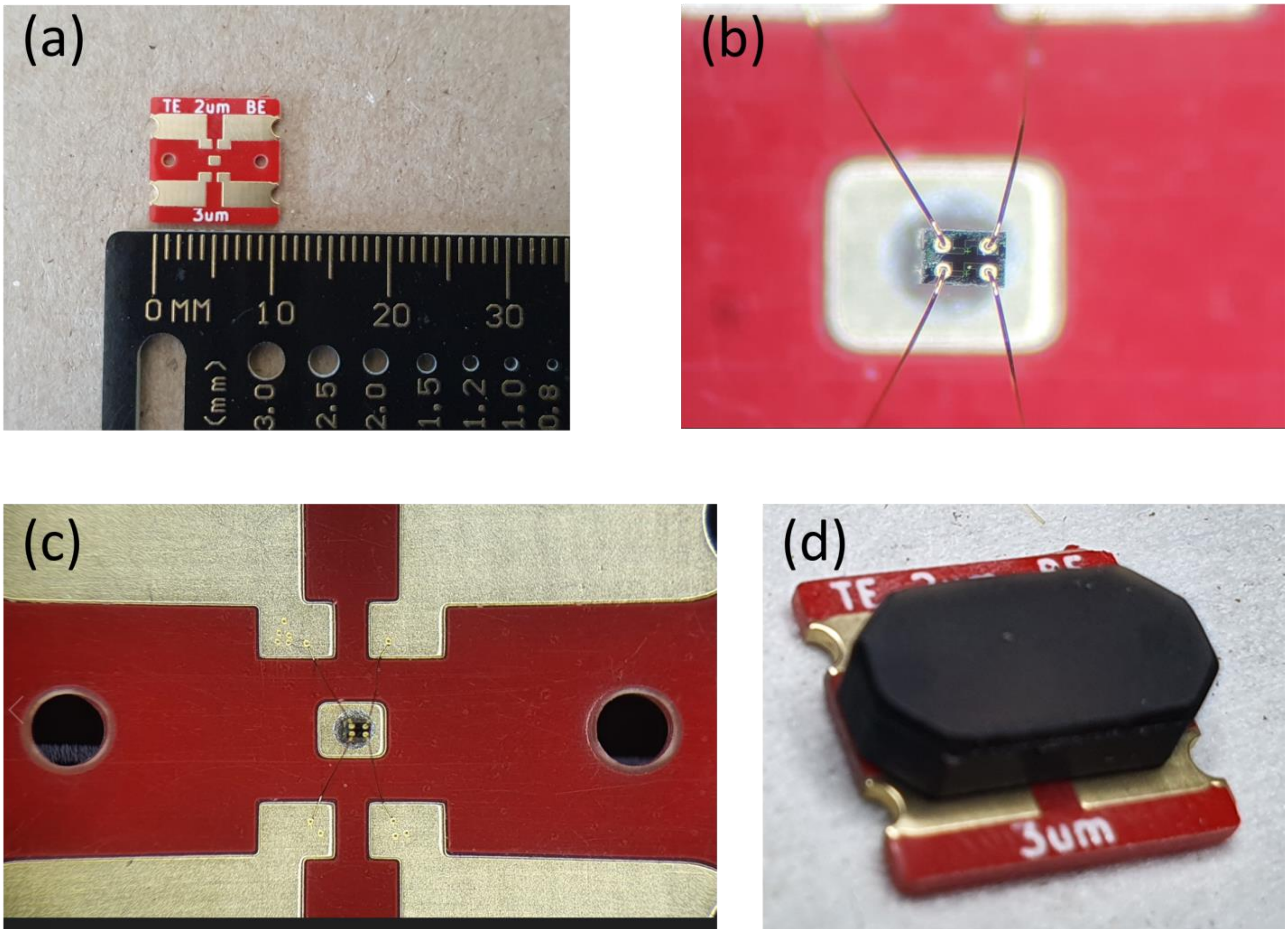

Fig. 6. (a) Bare PCB carrier. (b) Memristor attached with glue (silver epoxy adhesive) to the PCB metal pad. (c) Memristor wiring to the PCB. (d) Plastic enclosure protecting the bonded memristor.

### *D. Resetting the Memristor*

Once the memristor switches to an LRS (due to high input current signal), the TIA sensitivity is reduced. The memristor must be reset back to a HRS to allow the TIA to recover its sensitivity. A RESET pulse is applied to the memristor by the reset pulse control (see Figure 8). The external RESET signal is connected to the amplifier's inverting input through a 1kΩ resistor. The RESET signal is amplified by the operational amplifier (inverting configuration gain). The RESET signal amplification gain is, therefore,

$$RESET\ signal\ gain = -\frac{R_{mem}}{1k}\,. \tag{7}$$

A negative RESET signal is applied to switch the memristor back to an HRS.

## V. Circuit Evaluation

The proposed circuit was implemented on a PCB, which includes the OTA, TIA stage, and circuits for monitoring and resetting the memristor. The memristor's carrier board (Figure 6) was soldered to the PCB adjacent to the TIA amplifier. The evaluation board is shown in Figure 10. Prior to constructing the circuit, we conducted comprehensive simulations using LTspice. These simulations played a crucial role in optimizing the circuit and assessing the impact of memristor switching on the behavior of the TIA.

The evaluation board was connected to a vector network analyzer (VNA) to measure the frequency response when the memristor is in HRS and LRS. These measurements are shown in Figure 11. The bandwidth (although limited by the OTA and the low pass filter) is higher when the memristor is in the LRS. This behavior is similar to that of a TIA that has a low feedback resistor value (this can be explained by (1)). The difference in decibels between the TIA's gain in the LRS and HRS, as depicted in this figure, corresponds to the augmentation of the dynamic range demonstrated by our solution. Figure

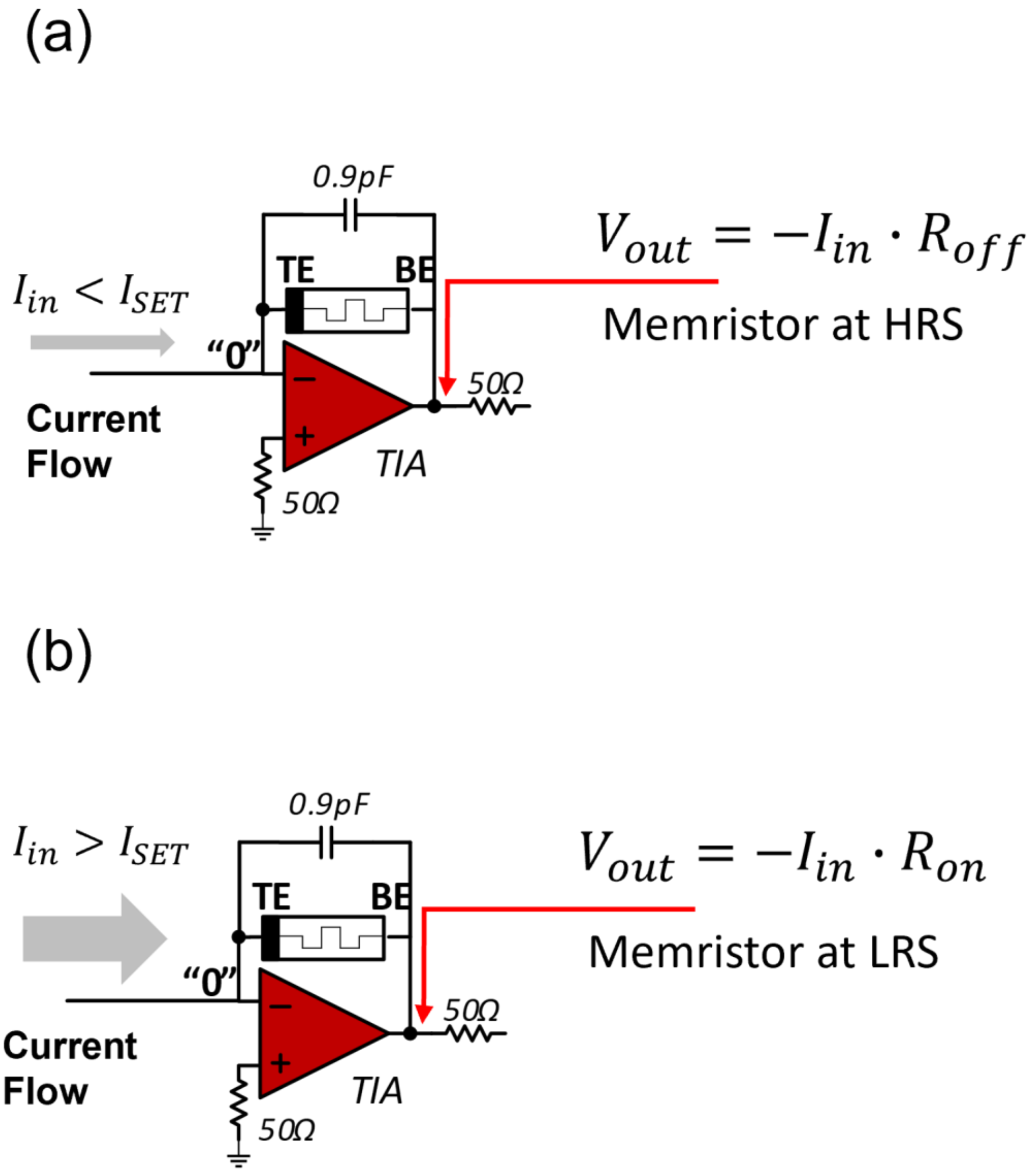

Fig. 7. Operation of the memristor-based TIA. (a) The memristor is initiated at HRS and TIA has high sensitivity for low input signal acquisition. (b) The memristor switches to LRS once the input current is higher than the threshold current. High input currents do not saturate the TIA in this state.

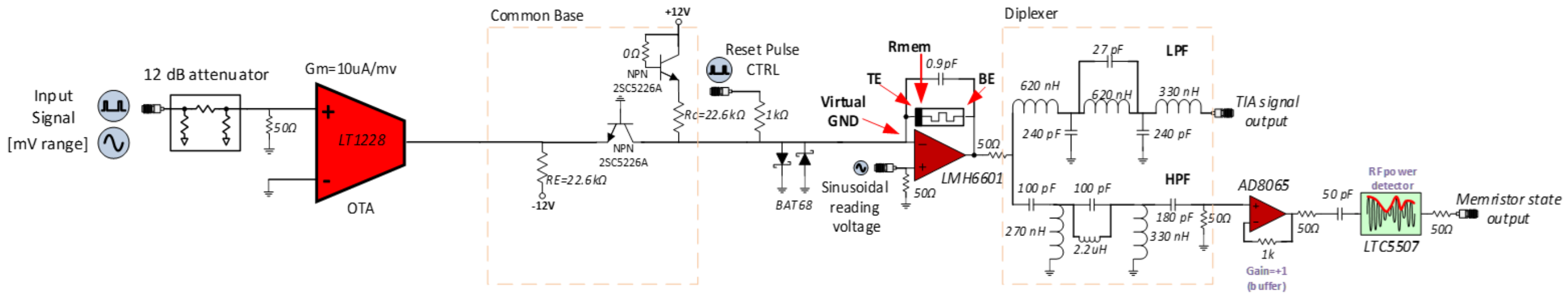

Fig. 8. Detailed schematic of the proposed memristor-based transimpedance AGC amplifier circuit and its testing setup. The output current of the OTA stage flows through the common-base and the TIA stage. An output diplexer after the TIA splits the signal to two paths, a low-pass path for the output signal and a high-pass path for the sinusoidal reading signal.

12 shows the switching of the memristor to LRS and the reduction of the TIA gain. The memristor was initialized to a HRS, and the input signal was configured to be a triangular wave. The change in the resistance is also observed in the RF detector output level waveform that had its amplitude reduced (after some settling time). The SET transition happens very fast (on a nano-second scale [35]) and, therefore, allows the TIA gain to decrease immediately when large signals appear. The switching of the memristor between its initial HRS to an LRS enables the dynamic range to be increased by the memristor's $\frac{R_{off}}{R_{on}}$ ratio. The memristor used for evaluation has a $\frac{R_{off}}{R_{on}}$ ratio of approximately 100 (although it varies from cycle to cycle) and, therefore, enabled the TIA dynamic range to be increased by up to 40dB. Using memristors with a higher $\frac{R_{off}}{R_{on}}$ ratio would further increase the dynamic range. By utilizing the memristor in the feedback loop of the TIA rather than a conventional resistor, the input-referred current noise of the TIA remains unaffected. Instead, this configuration solely enables an augmentation in the maximum input current into the TIA, following the memristor's transition to the Low-Resistance State (LRS). As a result, this approach facilitates a 40dB increase in the dynamic range, as indicated by (2). When a fixed resistor was employed in the TIA feedback, our circuit was found to have a dynamic range of 86.3 dB. However, upon integrating the memristor into the TIA feedback circuitry, the dynamic range significantly increased by 40 dB, reaching a total of 126.3 dB.

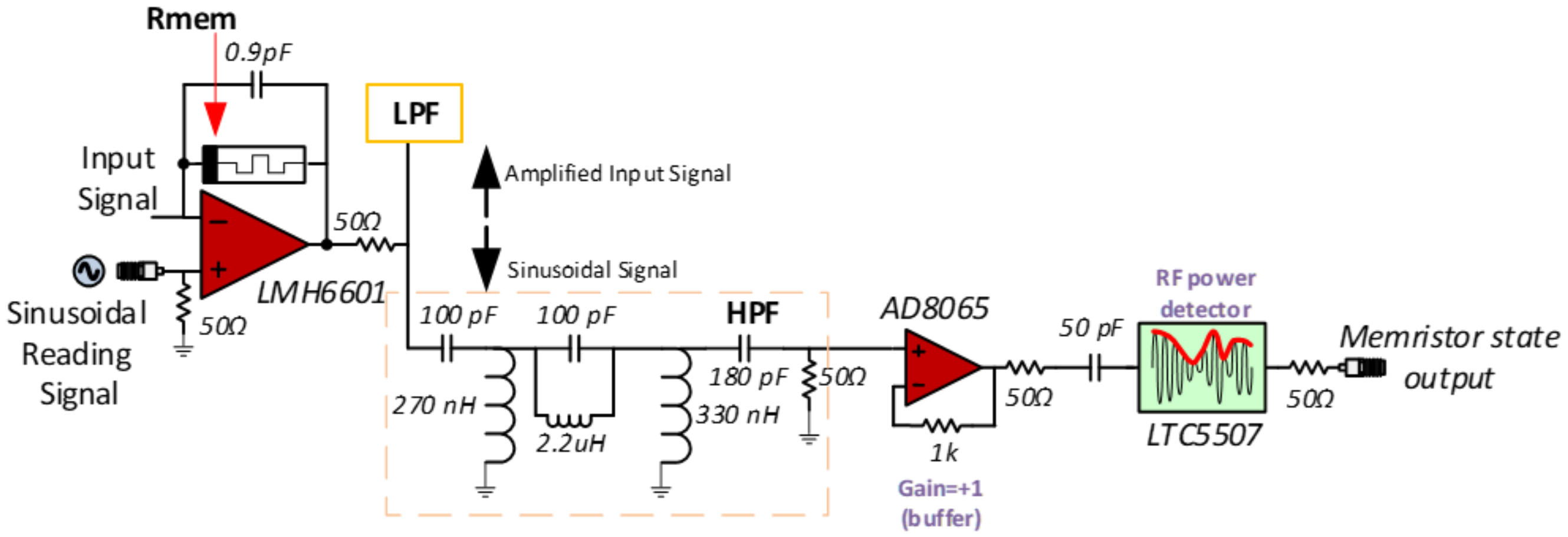

Fig. 9. TIA gain monitoring circuit. A sinusoidal source is connected to he non-inverting input of the amplifier. The sine signal is amplified according to the memristor's resistance value then undergoes filtering via the high-pass filter before being converted to a DC level by the RF power detector.

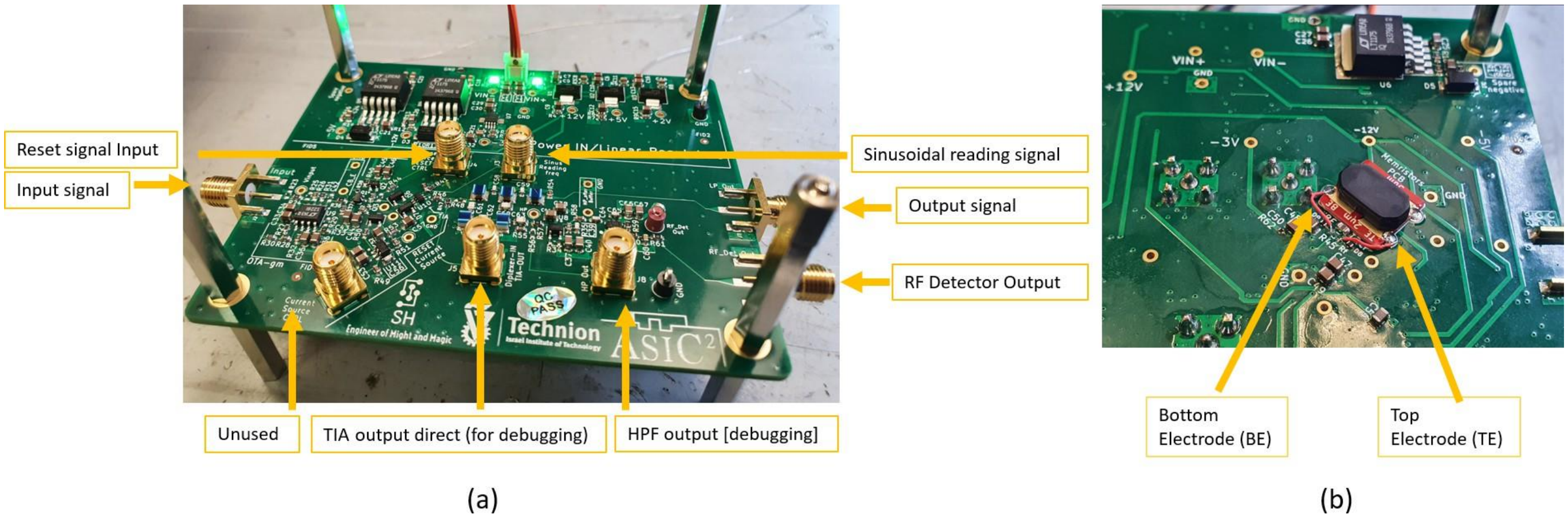

Fig. 10. TIA-based memristor printed circuit board (PCB). (a) Top and (b) bottom views.

A limitation observed during the evaluation was the cycle-to-cycle variations of the memristor. The SET threshold voltage exhibited approximately 20% cycle-to-cycle variation, while the $R_{off}$ had a high spread primarily due to the parameters of the RESET pulse used. Using memristors with lower $R_{off}$ variance is therefore desired and methods such as incorporating nanocrystals into the memristor oxide [36], [37], may improve this issue. Using circuit-level solutions can also improve immunity to cycle-to-cycle variations. For example, a better feedback network can be designed to reduce the effect of $R_{off}$ variance on the TIA performance. The memristor can be connected in parallel to another fixed-value resistor. As long as the memristor's $R_{off}$ is higher than the fixed resistor value, the variance of the $R_{off}$ would have a minor effect, as the fixed resistor would define the TIA's gain. Only once the memristor switches to an LRS, it would significantly lower the TIA gain and prevent saturation of the amplifier. Using this method, the dynamic range will be determined by the ratio of the fixed resistor value to the memristor's $R_{on}$.

Our design requires monitoring of the memristor state to reset it with an appropriate signal when the TIA needs to be restored to its high gain configuration for high sensitivity. Figure 13 shows the RESET signal. Before the pulse was activated, the memristor was in an LRS, and the TIA gain was low (evident by the low output voltage). After the successful RESET, the memristor switched back to its initial resistance, and the TIA gain was substantially increased (evident by the noisy output

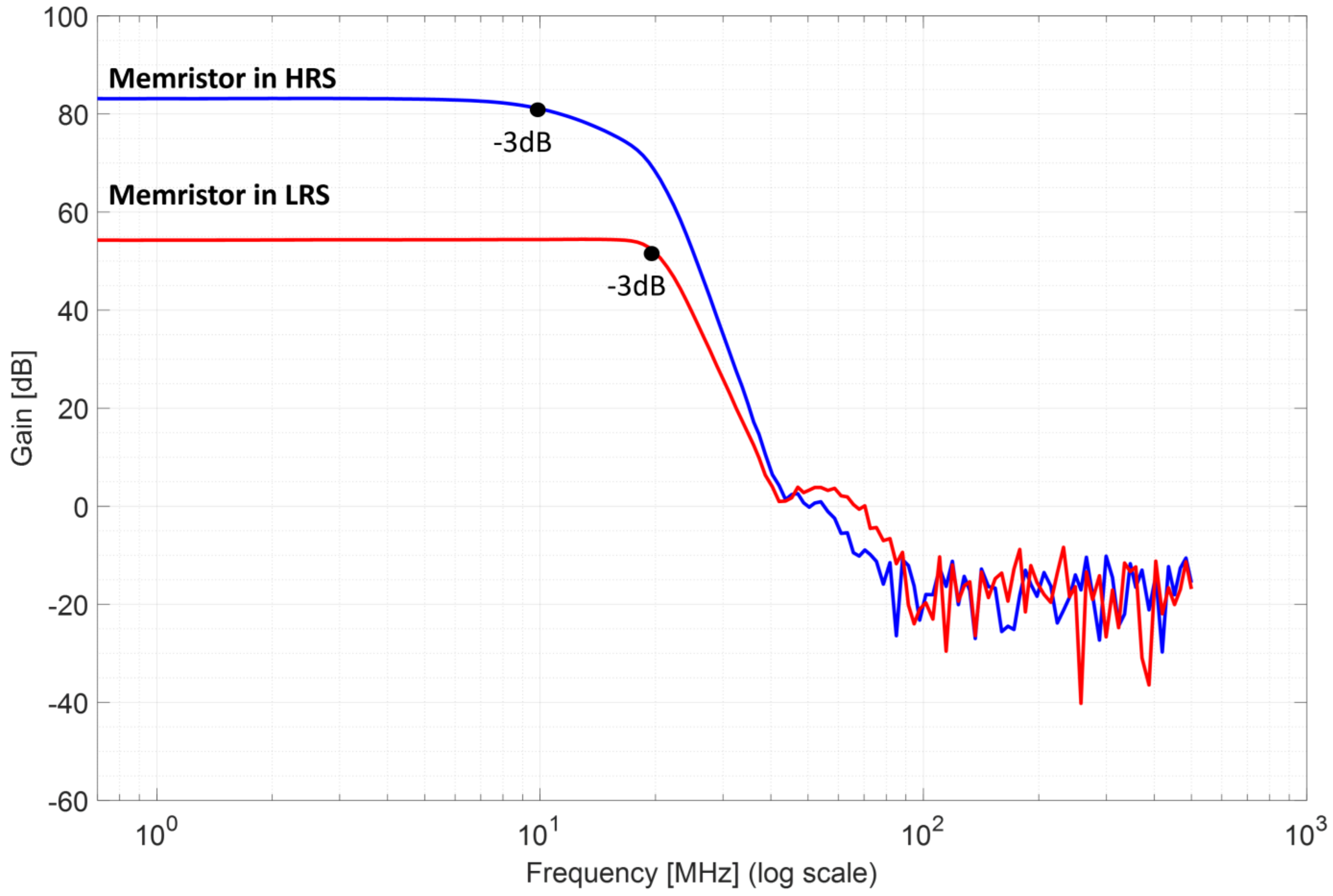

Fig. 11. Memristor-based TIA frequency response when the memristor is in HRS and LRS. In the LRS of the memristor, the bandwidth reaches 20 MHz, whereas in the HRS, the bandwidth is reduced to 11 MHz. The difference in TIA gain at low frequencies between the high-resistance state (HRS) and low-resistance state (LRS) corresponds to the expansion of the dynamic range facilitated by the proposed TIA.

voltage caused by environmental noise greatly amplified by the amplifier).

Since the required RESET pulse duration varies from cycle to cycle and is affected by the specific $R_{on}$ value, it was impossible to inject a RESET signal with the exact duration to cause the RESET transition. The RESET signal duration was always longer and continued even after the memristor switched to an HRS. Since the RESET signal appears at the TIA output after amplification that depends on the memristor resistance state as in (7), once the memristor switches to HRS, the RESET signal is tremendously amplified. It may cause the TIA to be saturated after a RESET signal is activated. This issue motivates the use of volatile memristors, which exhibits a short-term change in resistance in response to the applied voltage, rather than performing an explicit RESET operation.

Unlike conventional resistors, memristors exhibit Random Telegraph Noise (RTN) [38], which stems from fluctuations in the read current when voltage is applied to the memristor. This effect is particularly notable in memory applications, where memristors are employed. When memristors are in the HRS and their state is being read, the read current can experience significant fluctuations, potentially resulting in reading errors. These fluctuations occur due to electron traps within or near the ReRAM filament [38]. However, RTN should have minimal impact on memristor-based TIA circuits, primarily because the current is driven through the memristor in TIAs and not generated by it. Therefore, current fluctuations caused by electron traps within the filament should not affect the TIA's performance. Additionally, the RTN phenomenon operates on relatively long time scales, typically in the order of seconds, leading to low-frequency noise components. Consequently, the limited rise of noise at such low frequencies (several Hz) has a negligible impact on the overall noise characteristics of the TIA, particularly in cases where the TIA's bandwidth operates in the MHz and GHz range. Thus, the effect on the TIA's dynamic range is minimal.

Although this approach is similar to a logarithmic TIA [18], it demonstrates superiority in several aspects. Within the linear region, our solution achieves reduced total noise. Logarithmic TIA tends to have higher noise levels due to the inclusion of diode devices in the feedback circuitry. Additionally, our design exhibits a more rapid and distinct transition. Furthermore, the absence of active devices within the feedback network enhances the amplifier's stability of our approach.

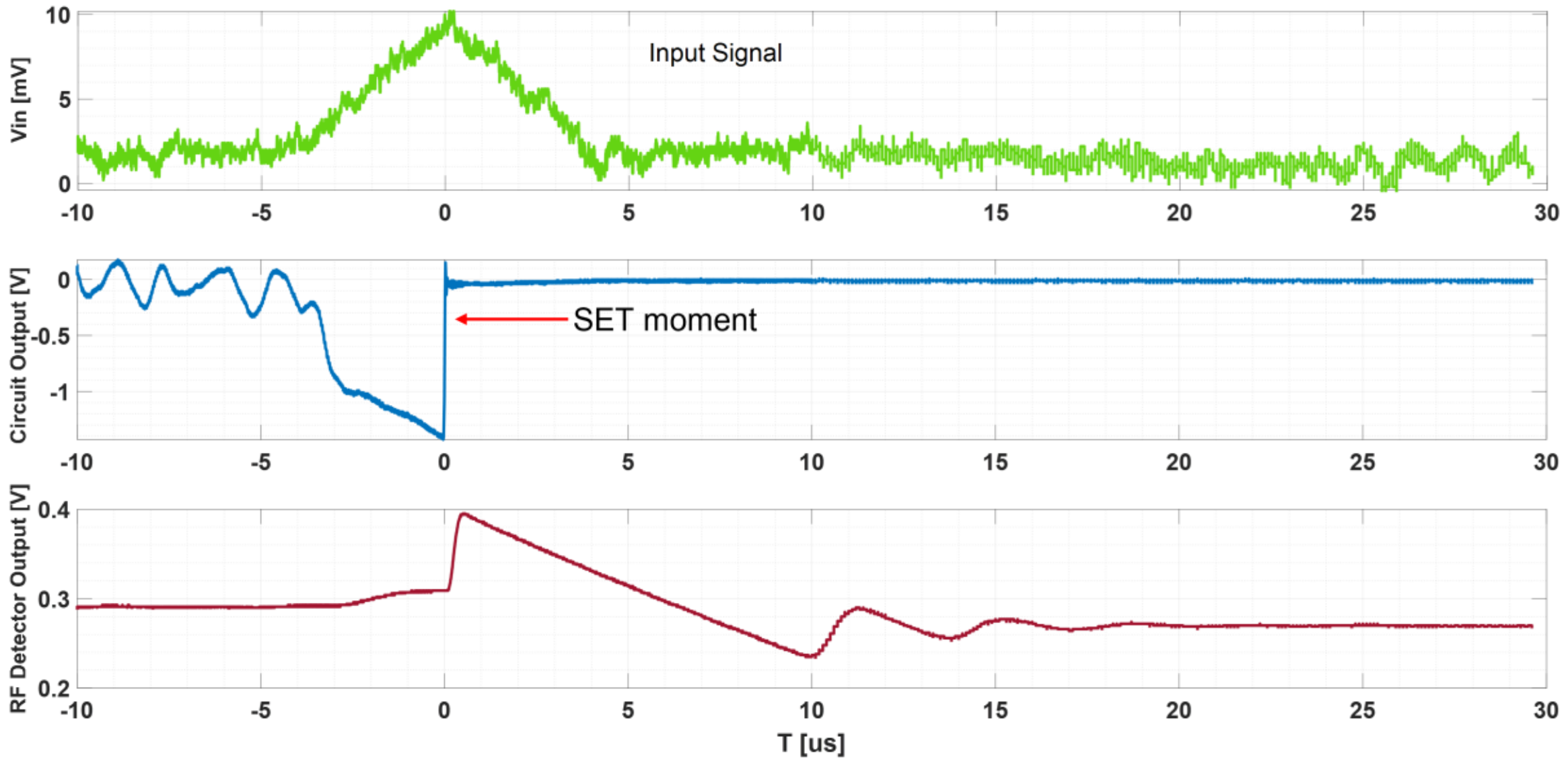

Fig. 12. SET transition of the memristor-based TIA. Once the input signal to the OTA reaches a certain threshold, the memristor switches to LRS and lowers the TIA gain. The measured transition time was found to be less than 10 nanoseconds. RF detector output level also indicates the change in the memristor state.

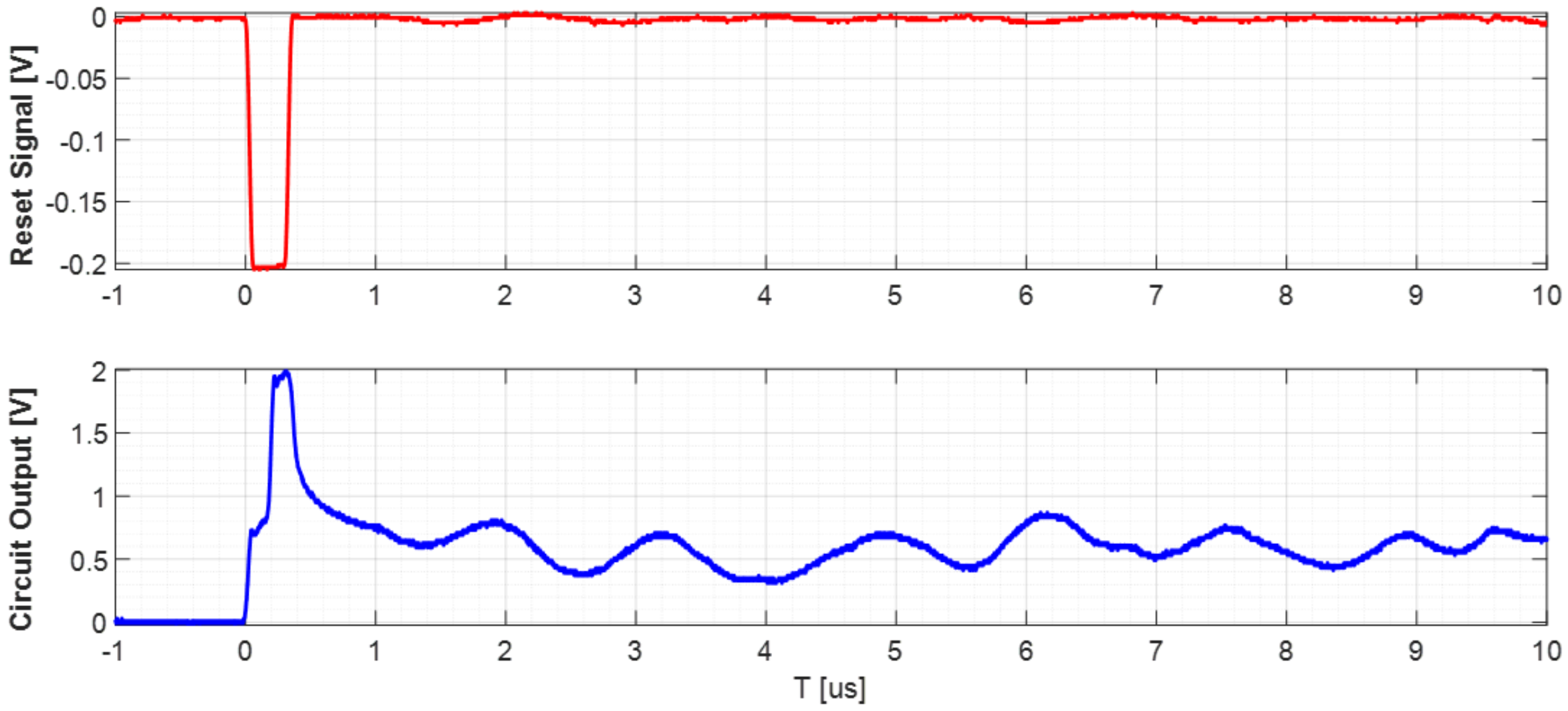

Fig. 13. Memristor's RESET signal. After the RESET, the memristor switched to HRS and increased the TIA gain. The highly amplified environmental noise indicates that TIA gain is higher after the RESET pulse.

## VI. Performance Comparison

We assessed our solution through a discrete PCB-based implementation. To facilitate a comparison with prior work, we designed the proposed TIA as an integrated circuit using the TSMC 0.18 $\mu$m process. This TIA design adopts the CMOS inverter topology outlined in [1]. Figure 14 illustrates the simulated schematic. The main performance parameters of the proposed memristor-based TIA, compared to state-of-the-art other wide dynamic range TIAs are summarized in Table I. For fair comparison, we have used the following figure-of-merit (FOM) [2]:

$$FOM = \frac{\sqrt{BW[GHz]}R_T[\Omega]C_T[pF]}{Noise[pA/\sqrt{Hz}]P[mW]}, \tag{8}$$

where BW represents the TIA bandwidth, $R_T$ stands for the TIA's transimpedance (gain), $C_T$ denotes the TIA's total input capacitance, the noise referred to by the FOM corresponds to the input-referred current noise density of the TIA, and $P$ represents the power consumption of the TIA.

Our proposed design enables the attainment of a substantial dynamic range, primarily governed by the memristor's $\frac{R_{off}}{R_{on}}$ ratio. By employing a memristor with a higher ratio, we could effortlessly push the boundaries of our TIA's dynamic range even further. Furthermore, our proposed solution demonstrates an exceptionally high FOM due to the simplicity inherent in our design, courtesy of the memristor. Since the memristor replaces the entire AGC circuit, the TIA design can be significantly streamlined, substantially reducing noise and lowering power consumption.

Our solution operates independently of other noise reduction, bandwidth expansion, or dynamic range enhancement techniques. As a result, it can be seamlessly integrated with these methods to achieve even higher dynamic range. Additionally, because memristors are integrated into back-end-of-line in the metal layers, their inclusion in the integrated circuit design does not necessitate an increase in the required area. The solution that surpasses our solution in terms of dynamic range (130 dB in [7] versus 113 dB in our solution) comes at the cost of exceedingly high power consumption, resulting in a notably low FOM. Solutions with a high FOM did not exhibit a similarly elevated dynamic range as our solution. This is because, with conventional methods (excluding the use of a memristor), it is challenging to simultaneously achieve low noise, high bandwidth, high gain, and a substantial dynamic range.

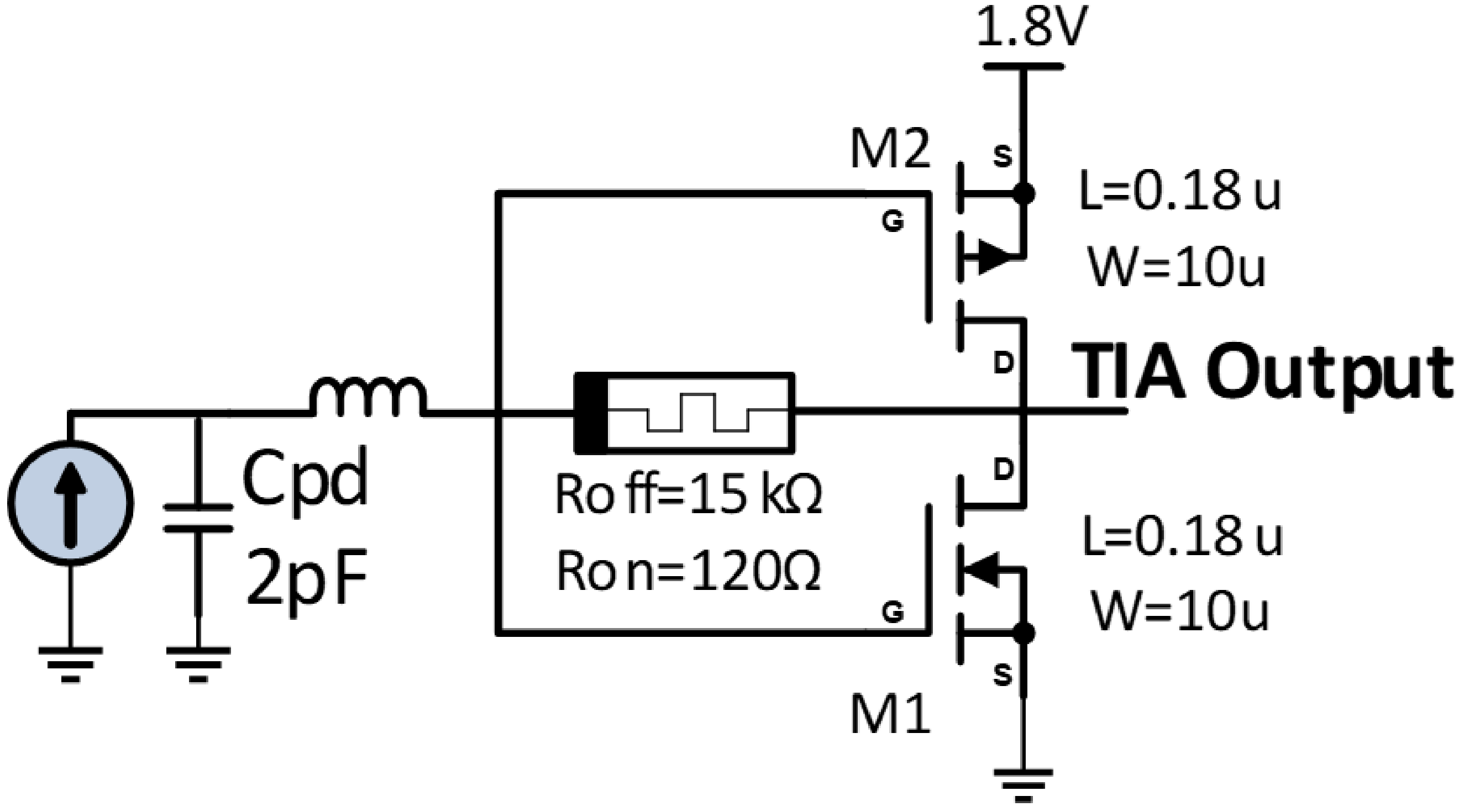

Fig. 14. Memristor-based transimpedance amplifier integrated circuit in CMOS inverter topology for 0.18$\mu$m process. Photodetector capacitance is 2pF.

## VII. Conclusion

High dynamic range is a crucial parameter in TIA circuits. Today's traditional high dynamic range circuits are highly complex and have an inherent tradeoff between bandwidth, dynamic range, and linearity. The proposed TIA shows that using the memristor as the feedback element in the TIA amplifier enables an AGC mechanism without the need for active feedback circuitry. This eliminates the requirement for monitoring the TIA output and adjusting the gain, which is necessary in many traditional solutions. We demonstrated our solution on a PCB and showed that the dynamic range of the TIA is extended proportionally to the memristor's $\frac{R_{off}}{R_{on}}$ ratio. The proposed design is also compared to prior work in integrated circuit (IC) design. Integrating memristors into IC design has additional benefits, as they are embedded in the metal layers and therefore do not necessitate an expansion of the amplifier's physical footprint or area.

Future advancements in our proposed solution involve investigating emerging memristive technologies capable of obviating the requirement for external monitoring and resetting circuits. This would also render the Hi/Lo diplexer unnecessary, leading to a reduction in the need for inductors and subsequently simplifying the circuit further. Another avenue of research involves refining the SPICE modeling of the memristor to achieve a more precise simulation of the RESET transition and cycle-to-cycle variations. Additionally, our ongoing investigation will aim to expand the versatility of our solution by accommodating more than two gain states, aligning it with other Automatic Gain Control (AGC) solutions that offer multiple steps.

TABLE I
PERFORMANCE COMPARISON OF STATE-OF-THE-ART TIA CIRCUITS WITH HIGH DYNAMIC RANGE

| | Input Dynamic Range (dB) | Gain Adjustment | $C_{pd}$ $(pF)$ | Trans -impedance Gain (dBΩ) | -3dB Bandwidth (MHz) | Input Referred Noise $(A/\sqrt{Hz})$ | Power Consumption (mW) | FOM | Technology |
|---|---|---|---|---|---|---|---|---|---|
| This Work | 113 | Memrisor -Based AGC | 2 | 83.5 | 98 | 2.1p | 1.1 | 4065 | CMOS 0.18$\mu$m |
| [7] | 130 | Compression | 2 | 89.5 | 250 | 58n | 330 | 0.00155 | BiCMOS 0.35$\mu$m |
| [39] | 97 | Compression | 4 | 59 | 227 | 8.96p | 3.45 | 55 | CMOS 0.13$\mu$m |
| [40] | 94 | Compression | 1 | 121 | 230 | 4.61p | 155 | 805 | CMOS 0.35$\mu$m |
| [41] | 85 | Compression | 15 | 68 | 280 | 13.14p | 8.7 | 174 | CMOS 40nm |
| [42] | 70 | VGCG with Post Amplification | 1.2 | 100.4 | 1260 | 3.81p | 39 | 945 | Dual Gate MOSFET |
| [17] | 69 | AGC | 0.35 | 80 | 1570 | 3.6p | 158.4 | 6.86 | Digital CMOS 0.11$\mu$m |
| [43] | 66 | Digital Code | 2 | 100 | 110 | 2.21p | 8 | 3751 | CMOS 0.18$\mu$m |

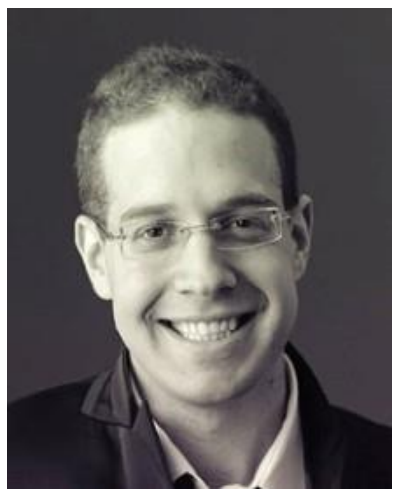

**Sariel Hodisan** (Student Member, IEEE) is an M.Sc. student at the Viterbi Faculty of Electrical and Computer Engineering, Technion. Sariel received his B.Sc. degree in Electrical Engineering and MSEE degree in Systems Engineering in 2009 and 2014, respectively, from the Technion. From 2009 to 2016 he worked in a leading defense organizations as a hardware and RF engineer. From 2016 through to today, he has been and still is a analog circuit engineer at Applied Materials. His current research interests include usage of memristors and new emerging devices for analog, high-speed and RF circuits.

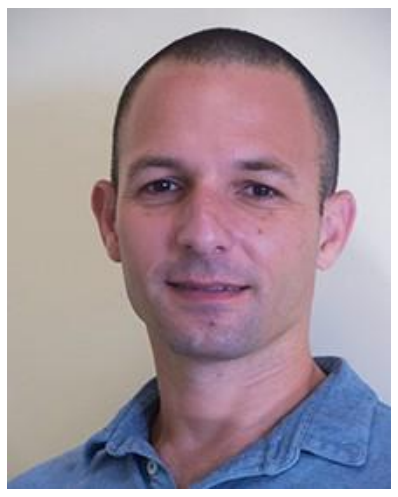

**Shahar Kvatinsky** (Senior Member, IEEE) is a Full Professor at the Andrew and Erna Viterbi Faculty of Electrical and Computer Engineering, Technion – Israel Institute of Technology. Shahar received the B.Sc. degree in Computer Engineering and Applied Physics and an MBA degree in 2009 and 2010, respectively, both from the Hebrew University of Jerusalem, and the Ph.D. degree in Electrical Engineering from the Technion – Israel Institute of Technology in 2014. From 2006 to 2009, he worked as a circuit designer at Intel. From 2014 to 2015, he was a post-doctoral research fellow at Stanford University. Kvatinsky is a member of the Israel Young Academy. He is the head of the Architecture and Circuits Research Center at the Technion, chair of the IEEE Circuits and Systems in Israel, and an editor of Microelectronics Journal and Array. Kvatinsky has been the recipient of numerous awards: the 2023 Uzi and Michal Halevy Award for Innovative Applied Engineering, the 2021 Norman Seiden Prize for Academic Excellence, the 2020 MDPI Electronics Young Investigator Award, the 2019 Wolf Foundation's Krill Prize for Excellence in Scientific Research, the 2015 IEEE Guillemin-Cauer Best Paper Award, the 2015 Best Paper of Computer Architecture Letters, Viterbi Fellowship, Jacobs Fellowship, an ERC starting grant, the 2017 Pazy Memorial Award, 2014, 2017 and 2021 Hershel Rich Technion Innovation Awards, the 2013 Sanford Kaplan Prize for Creative Management in High Tech, 2010 Benin prize, and seven Technion excellence teaching awards. His current research is focused on circuits and architectures with emerging memory technologies and the design of energy-efficient architectures.